\documentstyle[epsfig,12pt]{article}
\begin{document}
\begin{titlepage}
\vspace{-2mm} \rightline{hep-th/0006164} \vskip 1cm
\centerline{\Large {\bf Open and Closed Universes in Brane-World
Cosmology
 }} \vspace{1.5mm} \centerline{\Large
{\bf }} \vspace{10mm} \centerline{ Hyeong-Chan Kim,\footnote{
E-mail: hckim@theory.yonsei.ac.kr} Jae Hyung Yee\footnote{E-mail:
jhyee@phya.yonsei.ac.kr } and Mu-In Park\footnote{E-mail:
mipark@theory.yonsei.ac.kr } }

\centerline{{\it Department of Physics and Institute of Basic
Science, Yonsei University}} \centerline {{\it Seoul, 120-749,
Korea}} \vskip 10mm \centerline{\bf Abstract}

We find two types [called (S) and (A)] of new vacuum solutions of
open, flat, and closed universes which are inflating in the
brane-world scenario. We show that the warp factor of the
stabilized metric is universal for the three different kinds of
universes. For (S) type solution, we show that one
positive-tension brane universe solution is admitted as well as
two positive tension brane solution even if we consider the
vacuum solution. For (A) type solution, we find that the
inflating bulk solutions have black hole like regions and that
the full extended space is the R-S solution.
\end{titlepage}
\newpage

\section{{\large\bf Introduction}}

~~There has been considerable interest in the possibility of large
extra-dimension scenario and its applications to cosmology after
it was found that this scenario can resolve the long-lasting
problem of the hierarchy: The huge gap between the electroweak
scale $M_W$ and the Planck scale $M_P$, $M_P/M_W \sim 10 ^{16}$.
By allowing the extra dimensions to be {\it significantly larger}
than the Planck scale, one can try to relate the electroweak
scale to the fundamental higher-dimensional Planck scale, if the
matters are confined to our 4-dimensional
space-time~\cite{Rubakov,ADD,GRW}.

Randall and Sundrum [R-S]~\cite{RS1} suggested an alternative
solution to the hierarchy problem where the metric has exponential
dependence on the extra-dimension $y$:
\begin{eqnarray}
ds^2 = e^{-2\lambda r_c |y| }\eta_{\mu \nu} dx^\mu dx^\nu + r_c^2
dy^2 ,
\end{eqnarray}
where $y$  ranges $-1\leq y \leq 1$ and mirror symmetry for $y=0$
is assumed. This metric satisfies the vacuum Einstein equation
with the following relations between the cosmological constants
\begin{eqnarray}
\sqrt{\frac{-\Lambda}{24 M^3}}=\frac{\Lambda_{hid}}{24 M^3} =
-\frac{\Lambda_{vis}}{24 M^3} , \label{cosRel}
\end{eqnarray}
where $\Lambda$, $\Lambda_{hid}$, $\Lambda_{vis}$ are the
cosmological constants of the bulk, the hidden brane, and the
visible brane, respectively.
 In this model, the 4-dimensional Planck scale is given by
\begin{eqnarray}
M_P^2 = \frac{M^3}{\lambda} [1- e^{-2\lambda r_c} ]
\label{finetune}
\end{eqnarray}
and, for large $\lambda r_c $, it is $\lambda$ rather than $r_c$
that determines $M_P$. R-S argued that due to the warp factor
$e^{-\lambda r_c |y|}$  which has different values at the hidden
brane $y=0$ and at the visible brane $y=1$, any mass $m_0$ on the
visible brane corresponds to a physical mass $m= m_0 e^{-\lambda
r_c}$ and the moderate value $\lambda r_c \sim 37$ can produce
the huge hierarchy $M_P/M_W \sim 10 ^{16}$. Thus the gauge
hierarchy problem was converted to the problem of fixing the size
of the extra dimension. Moreover, this warp factor opened the
possibility of infinitely large extra-dimensions~\cite{RS2} due
to the localization of the gravity in addition to that of the
matter. This scenario is supported by the heterotic M-theory,
whose field theory limit is the 11-dimensional supergravity
compactified on $S_1/Z_2$ with supersymmetric Yang-Mills fields
living on two boundaries~\cite{HW,CS,K,V}.

On the other hand, R-S solution has two well-known defects: one is
that it contains a negative tension brane and the other is that it
needs a fine tuning~(\ref{finetune}) of the cosmological
constants of the bulk and of the branes.

For the first problem, it is known that this can be overcome by
permitting different cosmological constants~\cite{LR,DFGK} for
three spatial regions or by considering more complicated theories
with matter fields~\cite{KKOP,HBK}. But it seems that these
approaches are too artificial in our viewpoint: This is one of our
motivations.

In regard to the second problem, several
authors~\cite{Nihei,Kaloper,Kim2} have considered theories without
fine-tuning~(\ref{cosRel}) and they have found the inflationary
solutions. However, it was noticed~\cite{KLBDL} that the Hubble
parameter $H$ of these inflationary solutions in the brane-world
cosmology has a different behavior from that of the usual
4-dimensional Friedmann equation; in particular, the Hubble
parameter is proportional to the energy density on the brane
instead of the familiar dependence $H \sim \sqrt{\rho}$. Several
authors suggested that this problem can be
cured~\cite{Nihei,Kaloper,Kim2,CGKT,CGS} by requiring the
cancellation of the leading brane tension squared term
$\Lambda_{\rm brane}^2$ and the bulk cosmological constant
$\Lambda$ with matters on the brane. On the other hand, Kanti,
Kogan, Olive, and Pospelov~\cite{KKOP} derived sufficient
conditions which ensure a smooth transition to conventional
cosmology and Newton's law on the brane. The condition is the
existence of $T_{55} \propto (-\rho + 3 p)/L+ O(\rho^2)$, whose
value is responsible for the stabilization of the dilaton. We will
implicitly assume, in this paper, these types of resolution for
the inflating solution and we will not discuss it further. We
shall assume that there exists some stabilization mechanism, and
do not give a specific prescription.

Rather in this paper, we are concerned about the generalizations
of the (vacuum) inflating solutions to the spatially open and the
spatially closed universes, in contrast to the previous works
which have been restricted to the spatially flat universes. One
of the notable features of our new solutions is the existence of a
compact vacuum solution with one positive tension brane, and the
existence of the horizon on the bulk of the inflationary
solutions which has not been known yet.

In Sec. 2, we try to make a general effective action which
describes the brane cosmological systems. We use the $4+1$
splitting and the conformal transformation. We then choose a gauge
condition which should be satisfied by vacuum solutions. In Sec. 3
we generate several solutions which are inflationary. We find two
types of inflating solutions of the open, flat, and closed
universes with different warp factors: $\cosh^2(N\gamma|y|+\beta)$
and $\sinh^2(N\gamma|y|+\beta)$; we find that $\cosh^2$ type
solution allows one or two positive tension branes as well as the
one positive and one negative tension branes depending on the
initial data. We also give the coordinate transformations between
the R-S solution and our $\sinh^2$ type inflationary solution. It
is found that there exist an event horizon, which can be a past
horizon or a future horizon, but not both. Finally, we summarize
our results and present some comments in Sec. 4.

\section{{\large\bf Setting}}

~~We consider the vacuum solution of the brane world in D=5. The
fifth dimension $x^4=y$ satisfies reflection symmetry at $y=0$.
The five dimensional indices are denoted by $ M,N=0,1,\cdots,4$
and the four-dimensional indices of space-time by $a, b=0,1,2,3$.
The range of $y$ is restricted to $-1 \leq y\leq 1$.

 The 5-dimensional metric can be written, in general, as
\begin{eqnarray}
ds^2 =G_{AB} dx^A dx^B=e^{-2 \sigma}
   g_{ab}( dx^a+ N^a d y)(dx^b+ N^b d y)+ N^2 dy^2 ,
    \label{ds^2}
\end{eqnarray}
with $\sqrt{G}= N e^{-4 \sigma} \sqrt{g}$. The unit normal to the
time-like  section of $y=\mbox{constant}$  is denoted by
$n_A=(0,0,0,0,N)$ and $n^A = G^{A B}n_B=(- N^a/N, 1/N)$. The
induced metric on this  4-dimensional section is given by
\begin{eqnarray}
 h_{AB} = G_{AB} - n_A n_B .
\end{eqnarray}

From the Gauss-Codazzi relation, 5-dimensional curvature scalar
$^{(5)}R$ can be expressed by its 4-dimensional curvature $^{(4)}R
$ and the extrinsic curvature $\bar{K}_{AB}$ of the section up to
total derivatives:
\begin{eqnarray}
^{(5)}R = ^{(4)}\bar{R}+ \bar{K}^2 - \bar{K}_{AB} \bar{K}^{AB} +
\mbox{total derivatives}, \label{R5}
\end{eqnarray}
where $\bar{K}_{AB}= \frac{1}{2} {\cal L}_n G_{AB}$ is the Lie
derivative of the metric.

Using the conformal transformation $h_{ab} = e^{-2 \sigma}
g_{ab}$, one obtains $^{(4)}\bar{R}$:
\begin{eqnarray}
^{(4)}\bar{R} = e^{2 \sigma} \left\{ R + 6 g^{ab} \nabla_a
\nabla_b \sigma -6 \nabla_a \sigma \nabla ^a \sigma \right\} ,
\end{eqnarray}
where $R$ is the intrinsic curvature of $g_{ab}$.
 The extrinsic curvature is given by
\begin{eqnarray}
\bar{K}_{ab}= \frac{1}{2} {\cal L}_n h_{ab} = \frac{1}{2} {\cal
L}_n (e^{-2 \sigma} g_{ab} ) = - (n^C \partial_C \sigma) e^{-2
\sigma} g_{ab} + e^{-2 \sigma} K_{ab} ,
\end{eqnarray}
where $K_{ab}$ is the extrinsic curvature of $y=$ constant
section with metric $g_{ab}$.

Thus the Einstein action for 5-dimensional gravity including the
cosmological constant term can be written as
\begin{eqnarray} \label{S1}
S &=&  \frac{1}{16\pi}\int_M d^5x ~\sqrt{-G} \left[ M^3 R -
\frac{1}{2} \Lambda \right]
+ \sum_{\rm branes}S_{i}   \nonumber \\
  &=&\frac{M^3}{16 \pi} \int d^5 x \sqrt{g} e^{-2 \sigma} \left\{
   N[R+ 6 \nabla_a \sigma \nabla^a \sigma ] - 6 g^{ab} \nabla_a N
   \nabla_b \sigma  \right.  \\
   &+& \left. N e^{-2 \sigma} \left[ 12 (n^C\partial_C \sigma)^2 -
    6 K
   (n^C\partial_C \sigma) + K^2 - K^{ab}K_{ab} \right] -
   \frac{\Lambda}{2M^3} N e^{-2 \sigma}  \right\} + S_B ,
   \nonumber
\end{eqnarray}
where $S_{\rm i}$, $\Lambda$, and $M$ are the action for the
brane, cosmological constant of the bulk, and the fundamental
gravitational scale of the model, respectively. $S_B$ is the sum
of all the boundary terms in the canonical form of the action. The
brane action $S_i$ is given by
\begin{eqnarray}
S_{i} &=& -\frac{1}{32\pi}\int d^4 x~ e^{-4 \sigma} \sqrt{- g_i}
~\Lambda_{i} + S_{\rm matter} ,
   \label{SB}
\end{eqnarray}
and the variation of $S_i$ by,
\begin{eqnarray}
16\pi \delta S_i = \int d^5 x ~e^{- 4\sigma} \sqrt{-g} \Lambda_{i}
\delta(y-y_i) \left[
   2 \delta \sigma
   + \frac{1}{4} g_{ab} \delta g^{ab} \right]
   +16 \pi \delta S_{\rm matter} .
\end{eqnarray}
The extrinsic curvature $K_{ab}$ is explicitly written as
\begin{eqnarray}
K_{ab} = \frac{1}{2} {\cal L}_n g_{ab} = \frac{1}{2 N} \left[
\partial_y g_{ab} - \nabla_a N_b -\nabla_b N_a \right] .
\end{eqnarray}

We now consider a gauge choice that simplifies the action.  We
choose the metric $g_{ab}'$ to satisfy
\begin{eqnarray}
\partial_y g_{ab}'=\partial_y g_{ab} - \nabla _a N_b - \nabla_b
N_a  .  \label{Gauge1}
\end{eqnarray}
This choice of gauge can be justified because the gauge degree of
freedom  $x^a \longrightarrow x^a + \xi^a$ leads to
\begin{eqnarray}
g_{ab} ' = g_{ab} + \xi_{a;b} + \xi_{b;a} .
\end{eqnarray}
In this gauge~(\ref{Gauge1}), the shift function $N^a$ should
satisfy,
\begin{eqnarray}
\nabla _a N_b + \nabla_b N_a =0
\end{eqnarray}
which means that $N_a$ is a Killing vector of $g_{ab}$ or zero. If
$N_a$ is a Killing vector of $g_{ab}$, one can set $N_a$
proportional to a coordinate vector $e_a$. Then simple
reparametrization of $x^a$ coordinates can remove $N_a$. This
choice, $N_a=0$, can be justified because the spatial coordinates
can usually be adjusted to be orthogonal to $y$, but $N_t$ can not
be gauged away when there is energy exchange in $y$ direction. If
one is interested in the vacuum solutions as in this paper, $N_t$
can be trivially gauged away. We thus set $N_a=0$ from now on.

With this gauge choice the action becomes
\begin{eqnarray}
S &=&\frac{M^3}{16 \pi} \int d^5 x \sqrt{g} e^{-2 \sigma} \left\{
   N[R+ 6 \nabla_a \sigma \nabla^a \sigma ] - 6 g^{ab} \nabla_a N
   \nabla_b \sigma  \right. \nonumber \\
   &+&
    \frac{1}{N} e^{-2 \sigma}
      \left[ 12 (\partial_y \sigma)^2
    - 3 g^{ab} \partial_y g_{ab}(\partial_y \sigma)
     +\frac{1}{4}[(g^{ab}\partial_y g_{ab} )^2
           -g^{ac}g^{bd}\partial_y g_{ab} \partial_y g_{cd}  ]
      \right]  \\
   &-&\left.
   \frac{\Lambda}{2 M^3} N e^{-2 \sigma} \right\} + S_B .
   \nonumber
\end{eqnarray}
Note that there is a time-derivative term of $N$ in this action in
contrast to the usual time-slicing formulation. The variation of
this reduced action reproduces the Einstein equations, except for
the fact that the part of the Einstein equation, $G_{\mu 5} =0$,
which corresponds to the variation of $N_a$.

\section{{\large\bf Robertson-Walker models}}

~Since we are interested in the cosmological solutions, we assume
that the three-dimensional spatial section is homogeneous and
isotropic. The most general metric of this kind is given by
\begin{eqnarray}
g_{ab} dx^a dx^b = - d \tau^2 + a^2(\tau, y)  d \xi_{(3)}^2 ,
\end{eqnarray}
where $d \xi_{(3)}=\chi_{ij} dx^i dx^j $ is completely symmetric
spatial geometry. The intrinsic and the extrinsic curvature of
the metric are
\begin{eqnarray}
R = 6 \left[ \frac{\ddot{a}}{a} + \left(\frac{\dot{a}}{a}\right)^2
             \right] + \frac{6 k}{a^2} ,~~~~~K^a_{b} = \frac{1}{N}
{\rm diag}\left(0, \frac{a'}{a},\frac{a'}{a},\frac{a'}{a} \right)
,
\end{eqnarray}
where $k=1,0,-1$ for closed, flat, open spatial geometries,
respectively. The overdot ($\dot{a}$) denotes the time derivative
and the prime ($a'$) the derivative with respect to $y$.

We measure the scale of the universe by using the fundamental
scale of the system, $a(t,y)= e^r/M$, and then perform change of
variable $R= r-\sigma$. Then  the effective action reads
\begin{eqnarray}
S &=& \frac{6}{16 \pi} \int \sqrt{g} d^5 x N e^{3 R - \sigma}
   \left\{ - e^{2 \sigma} \left[ \dot{R}^2 + \frac{\dot{N}}{N}
   \dot{R} \right]  \right. \\
&+&\left. \frac{1}{ N^2} \left( R'^2 - \sigma ' R '\right)
 - \frac{\Lambda}{12 M^3} + k M^2 e^{-2 R} \right\}
   +S_B . \nonumber \label{SRW2}
\end{eqnarray}
The equations of motions are given by
\begin{eqnarray} \label{ddR}
&&\ddot{R} + 2 \dot{R}^2+ \dot{\sigma} \dot{R} -
  \frac{e^{-2\sigma}}{N^2} ( {R'}^2 - R ' \sigma ')
+ \left[-\frac{\Lambda}{12 M^3} + k M^2 e^{-2 R}
        \right]e^{-2 \sigma}=0 ,~~~~~ \\
&&\dot{R}^2 + \dot{R} \frac{\dot{N}}{N} + \frac{1}{N^2} e^{-2
\sigma} \left[ - 2 R'^2- \frac{ N '}{N} R '  -  R ''\right]
+\left(- \frac{\Lambda}{12 M^3} + k M^2 e^{- 2 R} \right)
e^{-2 \sigma} \nonumber \\
&&~~~~= \frac{e^{-2 \sigma}}{N} \sum_i
    \frac{\Lambda_i}{12 M^3}\delta(y-y_i)  , \\
&& 2 \ddot{R}+ 3 \dot{R} ^2 + 2 \dot{R}\dot{\sigma}+ 2
\dot{R}\frac{\dot{N}}{N} + \dot{\sigma} \frac{\dot{N}}{N}  +
\frac{\ddot{N}}{N} + \frac{1}{N^2} e^{-2 \sigma} \left[ -2 R ''
+\sigma ''  \right. \nonumber\\
&& \left. \frac{N'}{N} ( 2 R ' - \sigma ') - 3 R'^2  + 2 R '
\sigma ' -\sigma '^2- \frac{\Lambda N^2}{4 M^3} \right]+ kM^2
e^{-2 R}e^{-2 \sigma}
\\
&&~~~~ =
 \frac{e^{-2 \sigma}}{N}\sum_i
 \frac{\Lambda_i}{4 M^3} \delta(y-y_i) , \nonumber
\end{eqnarray}
where we have  used the fact that
\begin{eqnarray}
 \delta S_{i} = -\frac{6}{16 \pi} \int d^5 x \frac{\Lambda_i}{12 M^3} e^{3 R- \sigma}
 \delta(y-y_i) \left[ 3 \delta R - \delta \sigma \right] .
 \label{del SB}
\end{eqnarray}

If there exists a stabilization procedure~\cite{HBK}, $N$ will
become independent of $t$, and the $y$ dependence can be removed
by coordinate transformation. This leads to a constraint equation
which is independent of time derivatives if one sets
$N=\mbox{constant}$. This is the same kind of equation as the
Hamiltonian constraint, which implies the non-dynamical nature of
the lapse function. We then obtain
\begin{eqnarray}
R'' + {R'} ^2 - \frac{1}{3} \sigma '' + \frac{1}{3} {\sigma '}^2
+ \frac{\Lambda N^2}{18 M^3} = -\frac{N}{9} \frac{\Lambda_i}{M^3}
\delta(y-y_i)  \label{R''}
\end{eqnarray}
from Eqs.~(20), (21), and (22). Eq.~(24) implies an interesting
point: Since the equation is $k$ independent, the $y$-dependence
of the open, flat, and closed universe are the same if the $5$-th
direction is stabilized.

We now assume $\sigma'= - R'$, which corresponds to $a'=0$, and
solve Eq.~(24) at the bulk ($-1 < y < 1$). This leads to the two
sets of solutions:
\begin{eqnarray}
\sigma_{S,A}(y,t) &=& - \ln[e^{N \lambda (y-y_0(t))}
   \pm e^{-N\lambda(y-y_0(t))}]+ \sigma_0(t), \label{RW2}\\
R_{S,A}(y,t) &=&\ln[e^{N\lambda (y-y_0(t))}
        \pm e^{-N\lambda(y-y_0(t))} ]
        + R_0(t) , \nonumber
\end{eqnarray}
where  the subscripts $S$ and $A$ refer to $+,~ -$ sign,
respectively, and $\lambda^2 =\frac{-\Lambda }{24 M^3} $. Since
the dependence on $\sigma_0(t)$ can always be absorbed by
redefining the time coordinate, we set $\sigma_0(t)=0$.

By substituting both sets of Eqs.~(\ref{RW2}) to the first and
second equations of (\ref{ddR}) one obtains the following sets of
equations:
\begin{eqnarray}
\dot{y_0}(t) = 0, \label{y}\\
\dot{R_0}^2(t) -4 \lambda^2+ k M^2 e^{-2
R_0(t)} = 0  ,  \label{R0} \\
\ddot{R_0}(t) +N^2 \lambda^2 \dot{y_0}^2(t)  + 2 \dot{R_0}^2(t) -
8 \lambda^2 + k M^2 e^{-2 R_0(t)} = 0 \label{R02}.
\end{eqnarray}
Note that all of these equations (\ref{y}), (\ref{R0}),
and~(\ref{R02}) are not independent. Solving
Eqs.~(\ref{y},\ref{R0},\ref{R02}) gives the two types of metrics:
\begin{eqnarray}
 ds^2_S &=& \cosh^2 (\lambda N |y|-\beta) [ - d t ^2 + a^2(t)
  d \chi^2_{(3)}
 ] + N^2 dy^2    ,  \label{SBU} \\
 ds^2_A &=& \sinh^2 (\lambda N |y|+ \beta ') [ - d t ^2 + a^2(t) d
\chi^2_{(3)}
 ] + N^2 dy^2 ,~~~ \label{ABU}
\end{eqnarray}
where $a(t)= e^{R_0(t)}/M$ is given by
\begin{eqnarray}
a_{open}(t) = \frac{1}{\lambda} \sinh \lambda t, ~~ a_{flat}(t) =
\frac{1}{\lambda} e^{\lambda t}, ~~
 a_{closed}(t) =\frac{1}{\lambda} \cosh \lambda t,
\end{eqnarray}
for open, flat, and closed universe, respectively, and
$\displaystyle \lambda =\sqrt{\frac{-\Lambda}{24 M^3}} $, $\beta
= -N\lambda y_0$, and $\beta'= N\lambda y_0 (>0)$ are constants.

Let us first consider $\cosh$ (S) type solutions. The solutions
consist of two branes which are located at $y=0$ and at $y=1$. If
$\beta <0$, one of the two branes has positive tension and the
other has negative tension. On the other hand, if $N\lambda
> \beta > 0$, the two branes have positive tension. If
$\beta=0$, there exist one compact positive brane
solution
\begin{eqnarray} \label{1brane} ds^2 &=& \cosh^2
(N\lambda y) [ - d t ^2 + a^2(t)
  d \chi^2_{(3)}
 ] + N^2 dy^2    ,
\end{eqnarray}
in addition to the two brane solution, where one of the two branes
has zero tension. The brane in this solution~(\ref{1brane}) is
located at the boundary $y=1$ and this surface is identified with
that at $y=-1$. Due to the symmetry of the warp factor this
solution satisfies $S_1/Z_2$ boundary condition automatically.

We now consider the boundary condition:
\begin{eqnarray}
\left.R''\right|_{\mbox{at Brane}}= - \sum_i \frac{N^2
\Lambda_i}{12 M^3}  \delta(y-y_i) , \label{BC}
\end{eqnarray}
which implies the discontinuity of $R'$ at $y=y_i$. In
Eq.~(\ref{BC}) we have neglected terms related to $R'$ and $\sigma
'$ which are finite at $y=y_i$. The resulting brane tensions
satisfying this boundary condition are
\begin{eqnarray}
\lambda_0 = \frac{\Lambda_0}{24 M^3} = \lambda
    \tanh \beta , ~~ \lambda_1 =\frac{\Lambda_1}{24 M^3}
     =  \lambda
    \tanh\left(N \lambda - \beta \right) .
 \label{}
\end{eqnarray}
In the case of one brane solution, only $\lambda_1$ is meaningful.
Note that all the cosmological constants $\lambda_0$ and
$\lambda_1$ are smaller than $\lambda$ which is in contrast to the
$(A)$ type solution~[See below].

The Hubble parameters, $\displaystyle H= \frac{1}{ae^{-\sigma}}
\frac{d (a e^{- \sigma}) }{d\tau}$, where $\tau= e^{-\sigma} t$ is
the proper time, at each boundaries are
\begin{eqnarray} \label{}
H(0,t) = \sqrt{ \lambda^2- \lambda_0^2} ~( \coth \lambda t, 1 ,
\tanh \lambda t ), ~~ H(1,t) = \sqrt{ \lambda^2- \lambda_1^2} ~(
\coth \lambda t, 1 , \tanh \lambda t ) ,
\end{eqnarray}
for the open, flat, and closed universes, respectively.  The
Hubble parameters of the open and closed universe are dependent
on time, but asymptotically approach to the flat universe value
and the static limit $(H \rightarrow 0)$ corresponds to $\lambda_i
\rightarrow \lambda$~\cite{Kim2}. The current observation
restricts the Hubble parameter of the visible brane:
\begin{eqnarray} \label{Hubble0}
H(0) = \sqrt{\lambda^2- \lambda_0^2} < 10^{-60} M_P ,
\end{eqnarray}
where $\lambda = O(M_P)$ is assumed. Although Eq.~(\ref{Hubble0})
appears to require highly precise fine tuning, it actually is not
the case: Since the cosmological constant is related to the size
of extra-dimension, this constraint only implies $\beta \simeq 60
$.

Gauge hierarchy demands that the difference of the scale factors
between the hidden and visible branes is of the order of
$10^{16}$, which means $\lambda N- 2 \beta \simeq 37$. This
implies the size of the extra dimension,
\begin{eqnarray} \label{}
  N =\frac{1}{2\lambda}  \ln \left( \frac{\lambda +
\lambda_0}{\lambda -\lambda_0}~ \frac{\lambda +
\lambda_1}{\lambda-\lambda_1} \right) \simeq 157 ~l_p .
\end{eqnarray}
This size of extra-dimension is a reasonable one which also
guarantees the efficiency of semi-classical treatment of the
universe.

Now, let us consider the $\sinh$ (A) type solutions, which are
already discussed in Ref.~\cite{Kim2} for flat spatial geometry.
They admit two brane solutions with one positive and one negative
tension as usual. The Hubble parameters at each boundaries are
\begin{eqnarray} \label{}
H(0,t) = \sqrt{ \lambda_0^2- \lambda^2} ~( \coth \lambda t, 1 ,
\tanh \lambda t ), ~~ H(1,t) = \sqrt{ \lambda_0^2- \lambda^2} ~(
\coth \lambda t, 1 , \tanh \lambda t ) ,
\end{eqnarray}
for open, flat, and closed universe, respectively. In this case,
the size of the extra-dimension to solve the gauge hierarchy
problem,
\begin{eqnarray} \label{}
 N =\frac{1}{2\lambda}  \ln \left( \frac{-\lambda_0 -
\lambda}{\lambda_1 -\lambda}~ \frac{\lambda_1 +
\lambda}{-\lambda_0+\lambda} \right),
\end{eqnarray}
does not have a unique value, but varies depending on how
$\lambda_0$ and $\lambda_1$ approach  $\lambda$.

\begin{figure*}[tbh]
\vspace{0.5cm}
 \centerline{
\epsfig{figure=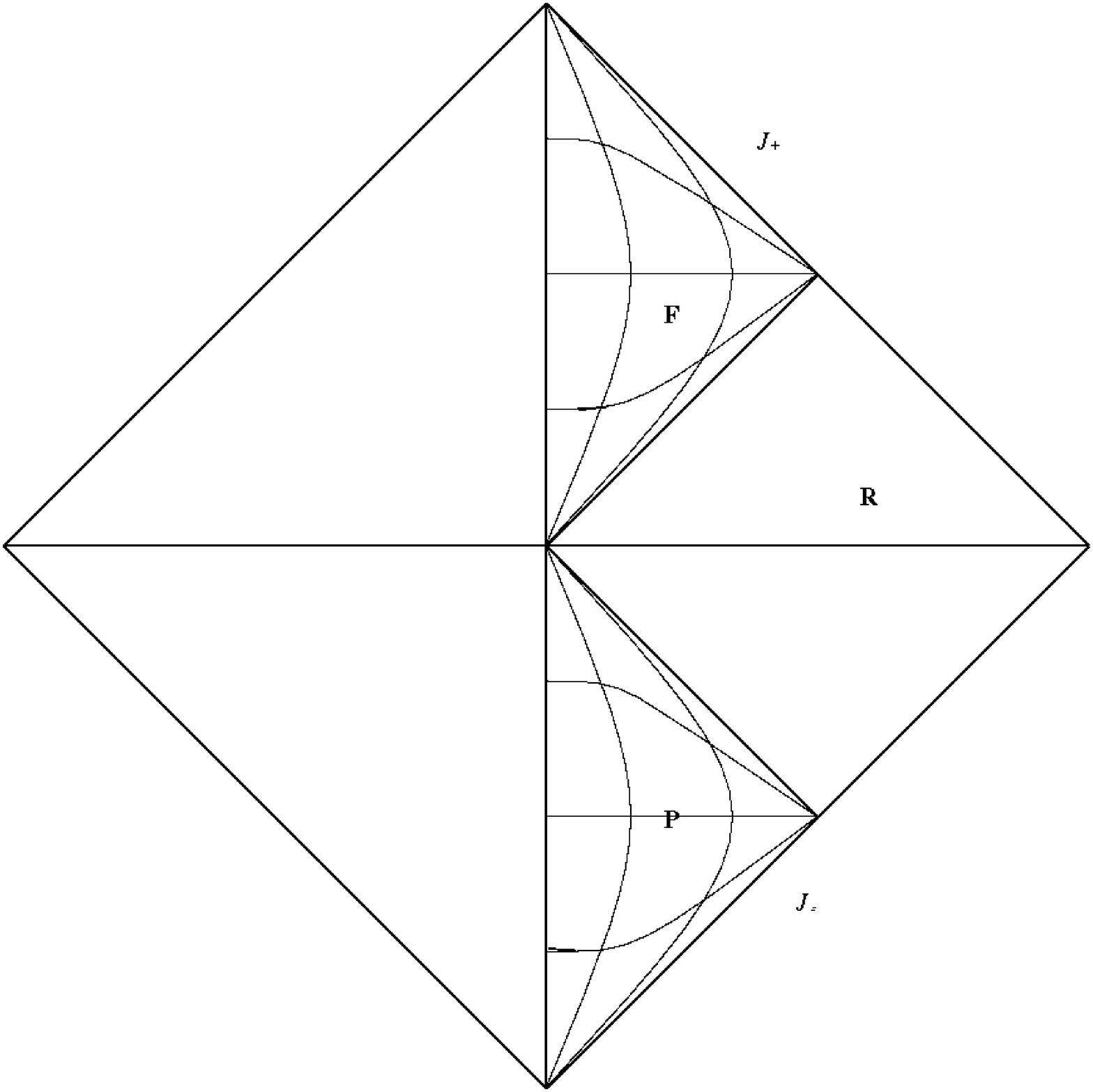,height=8cm}}
 {\footnotesize Fig. 1. Penrose diagram of the extended space. \\
Curves passing the origin represent $y$= constant. The brane
resides on this trajectory. $\bar{x}=0$ line represents infinity.
Inflating solution (\ref{ABU}) is confined region (P). The metric
confined in (F) is a deflating solution which can be obtained by
$t \rightarrow - t$ from Eq.~(\ref{ABU}). So an inflating
solution has a black hole like region, and a deflating solution
has a white hole like region. The horizons $\bar{u} =0$ or
$\bar{v}=0$ correspond to $y=0$.
 }
\end{figure*}

An interesting point of this solution, which has not been well
discussed yet, is the existence of a horizon at $y=0$ for $\beta
=0 $ case of the bulk solution~(\ref{ABU}). For convenience, we
set $N=1$. This solution should be extendible in $y$ coordinate.
We consider only the $y-t$ space to show the extendibility and we
consider only the flat space case for simplicity.

The metric of the fully extended space [See Fig. (1) for the
Penrose diagram.] is given by
\begin{eqnarray} \label{RSSol}
ds^2 = \frac{1}{\lambda^2} \frac{1}{\bar{x}^2} \left[ - d\bar{u}
       d \bar{v} +  d \chi_{(3)}^2 \right],
\end{eqnarray}
which is nothing but the Randall-Sundrum solution, as can be
expected from the fact that both the spaces (A) and the R-S
solution have the constant negative curvature. The coordinate
transformation between (\ref{RSSol}) and solution (A) is given by,
\begin{eqnarray}
 e^{\mp 2 \lambda t} = \bar{u} \bar{v}, ~~
 \cosh (N \lambda y) = \mp \frac{\bar{t}}{\bar{x}}  ,
\end{eqnarray}
where the negative sign corresponds to the inflating universe (P)
and the positive sign corresponds to the deflating universe (F).
The deflating universe solution can be obtained by setting $t
\rightarrow -t$ in Eq.~(\ref{ABU}). These inflating and deflating
solutions cover only the region $\bar{u}, ~\bar{v} < 0$ or
$\bar{u}, ~\bar{v} > 0$ of the whole space, respectively. The
inverse transformations are
\begin{eqnarray}
\bar{x} = {\rm cosech} (N \lambda y) ~e^{-\lambda t}, ~~ ~~
\bar{t} = \coth (N \lambda y) ~e^{-\lambda t} .
\end{eqnarray}

The metric of a localized observer in region (R) is given by
\begin{eqnarray} \label{R}
ds^2 = N^2 \sin^2 \lambda \tau ~dy^2 +
 e^{-2  N\lambda |y|} \left[ - e^{-2 N \lambda  |y|}~ d
 \tau^2  +  \frac{1}{\lambda^2} \sin^2 \lambda \tau ~
   d \chi_{(3)}^2 \right]  .
\end{eqnarray}
The coordinate transformation between Eqs. (\ref{RSSol}) and
(\ref{R}) is given by
\begin{eqnarray}
  - \bar{u} \bar{v} = e^{-2 \lambda N y} , ~~
    \frac{\bar{t}}{\bar{x}} = - \cos \lambda \tau  , ~~
 \label{}
\end{eqnarray}
where the ranges of $t$ and $y$ are $ 0 \leq \tau \leq
\pi/\lambda$ and $-\infty < y < \infty$, respectively. The scale
factor of the visible brane expands to $e^{-2 N \lambda}/\lambda$
and then recontract to zero within a finite time. The Hubble
parameter does not depends on $y$ coordinates and is
\begin{eqnarray}
H(t) = \lambda \cot \lambda \tau .
\end{eqnarray}
 If some brane universe lives in this space, its 5-th dimensional size
expands in time and then recontract. So this configuration does
not have stable brane universe solution.

This diagram [Fig.(1)] opens us another interesting possibility:
R-S brane as a wormhole. Consider two inflating brane worlds at
(P) and (F). These two branes with small positive tensions will
change background geometry somewhat, but we assume that this
change is small enough not to destroy its causal structure. Then
assume a third brane which is the R-S brane. R-S brane passes both
regions (P) and (F) and meets at some points with the two brane
worlds at (P) and (F). If one can move from one brane to another
brane when the two branes cross, the R-S brane can be used as a
wormhole between the two branes, which do not cross each other.

Some general remarks are in order. First, the limit, $\lambda
\rightarrow 0$ and large $\lambda N y$, of the open universe
solution for (S) is
\begin{eqnarray}
ds_{(S)}^2 = e^{-2 s |y|} \left[ - dt^2 + t^2 d \chi_{(3)}^2
\right] +  N^2 dy^2 .
\end{eqnarray}
The corresponding solution for the flat universe is the R-S one
and there is no solution for the closed universe. The cosmological
constant of each branes are given by Eq.~(\ref{cosRel}).

The second is that $05$ part of the Einstein equation, which comes
from the variation of $N_a$ in action~(\ref{S1}),
\begin{eqnarray}
R ' \frac{N'}{N} - \dot{R}{\sigma}' - \dot{R} R' - \dot{R}'=0 ,
\end{eqnarray}
is automatically satisfied with our solutions. This equation can
be used to obtain more general solutions than Eqs.~(\ref{ABU}) and
(\ref{SBU}): For example, by imposing only the stabilization
condition $N' =0$, one gets
\begin{eqnarray}
\sigma(t,y) + R(t,y) = - \ln \dot{R}(t,y) + f(t),
\end{eqnarray}
or $R'=-\sigma '$, instead, gives
\begin{eqnarray}
N(t,y) = R '(t,y) g(y),
\end{eqnarray}
which may allow more general solutions.

\section{\large \bf Summary and Discussion}

We have constructed the brane-world solutions of the
Robertson-Walker type by writing the effective action of a brane
world in $4+1$ splitting form. We have obtained spatially open,
flat, and closed universe solutions which are inflating in $t$ and
static in $y$ direction. There exist two types of solutions (S)
and (A). Notably, the one positive-tension brane universe
solution is admitted for (S) type as well as two brane solutions
with two positive tension branes or one positive and one negative
tension branes. We also estimate a reasonable size of the $5$-th
dimension which solves the cosmological constant and the $M_P/M_W$
hierarchy problems. Finally, in the case of (A) type solution, we
obtain its covering space which is nothing but the R-S solution.
Moreover, this (A) type solutions have a white-hole like region
or a black-hole like region.

We finally note that there is a minimal size for the closed
universe solution in Eqs.~(\ref{SBU}) and (\ref{ABU}). This size
is $\sinh(\beta)/\lambda$ or $\cosh(\beta)/\lambda$ and determined
by the cosmological constant of the bulk and the tension of the
branes:
\begin{eqnarray} \label{}
  L_{{\rm min},A} =\frac{\lambda_0}{\lambda}
    \frac{1}{\sqrt{\lambda^2-
         \lambda_0^2}}  , ~~~~
   L_{{\rm min},S} =\frac{1}{\sqrt{\lambda^2-
         \lambda_0^2}} .
\end{eqnarray}
These sizes are very large [$O(10^{60}~ l_P)$] if the
condition~(\ref{Hubble0}) is satisfied and it seems that the
closed universe is excluded as a candidate for our universe.
However, it is not evident whether or not the cosmological
constant satisfies Eq.~(\ref{Hubble0}) even at the initial epoch
of the universe. Moreover, if there is matter on the bulk at the
initial epoch, the contribution of matter will dominate $k$ term.
Then the possibility of the closed universe can not be excluded.

\centerline{{\large\bf Acknowledgments}}
\vspace{3mm}
 \noindent
H.C.K and M.I.P are supported by Korea Research Foundation Grant
KRF-99-005-D00010, D00011.

\newpage

\setcounter{equation}{0}


\begin{thebibliography}{[00]}

\bibitem{Rubakov} V. A. Rubakov and M. E. Shaposhnikov, Phys.
Lett. {\bf B 125}, 136 (1983).
\bibitem{ADD} N. Arkani-Hamed, S. Dimopoulos and G. Dvali, Phys.
Lett. {\bf B 429 }, 263  (1998), hep-ph/9803315.
\bibitem{GRW} G. F. Giudice, R. Rattazzi, and J. D. Wells, Nucl.
Phys. {\bf B 544}, 3 (1999), hep-ph/9811291.
\bibitem{RS1} L. Randall and R. Sundrum, Phys. Rev. Lett. {\bf
83}, 3370 (1999), hep-ph/9905221.
\bibitem{RS2} L. Randall and R. Sundrum, Phys. Rev. Lett. {\bf
83}, 4690 (1999), hep-th/9906064.
\bibitem{HW} P. Horava and E. Witten, Nucl. Phys. B460, 506
(1999).
\bibitem{CS} M. Cvetic and H. Soleng, Phys. Rep. {\bf 282}, 159
(1997).
\bibitem{K} A. Kehagias, Phys. Lett. B469, 123 (1999), hep-th/9906204.
\bibitem{V} H. Verlinde, {\it Holography and compactification},
hep-th/9906182.
\bibitem{LR} J. Lykken and L. Randall, hep-th/9908076.
\bibitem{DFGK} O. DeWolfe, D.Z. Freedman, S.S. Gubser, and A.
Karch, hep-th/9909134.
\bibitem{KKOP} P. Kanti, I.I. Kogan, K.A. Olive, and M. Pospelov,
Phys. Lett. {\bf B 468}, 31 (1999), hep-ph/9909481; Phys. Rev.
{\bf D 61 }, 106004 (2000), hep-ph/9912266; hep-ph/0005146 ; P.
Kanti, K. A. Olive, and M. Pospelov, {\it Solving the Hierarchy
Problem in Two-Brane Cosmological Models}, hep-ph/0005146.
\bibitem{HBK} H.B. Kim, Phys. Lett. {\bf B 478} 285 (2000),
 hep-th/0001209.
\bibitem{Nihei} N. Nihei, Phys. Lett. {\bf B 465} 81 (1999),
hep-ph/9905487.
\bibitem{Kaloper} N. Kaloper, Phys. Rev. {\bf D 60} 123506 (1999),
hep-th/9905210.
\bibitem{Kim2} Hang Bae Kim and Hyung Do Kim, Phys. Rev. {\bf D
61}, 064003 (2000), hep-th/9909053.
\bibitem{KLBDL} N. Kaloper
and A. Linde, Phys. Rev. {\bf D 59} (1999) 101303,  ;
P.Bin\'{e}tury, C. Deffayet, and D. Langlois, hep-th/9905012.
\bibitem{CGKT} C. Cs\'{a}Ki, M.
Graesser, C. Kolda, and J. Terning, Phys. Lett. {\bf B 462} 34
(1999), hep-ph/9906513.
\bibitem{CGS} J. M. Cline, C. Grojean, and G. Servant, Phys. Rev.
Lett. {\bf 83}, 4245 (1999), hep-ph/9906523.





\end{thebibliography}
\end{document}